\documentclass[aps,preprint,eqsecnum]{revtex4}
\usepackage{epsfig}
\begin{document}
\newcommand{\be}{\begin{eqnarray}}   
\newcommand{\ee}{\end{eqnarray}}
\newcommand{\ep}{\epsilon}
\newcommand{\si}{\sigma}

\preprint{DO-TH 04/09}
\vspace*{1cm}
\title
{Testing the Equivalent Photon Approximation of the Proton in the Process 
{\mbox{\boldmath $e p\rightarrow \nu W X$}}}
\author{\bf Cristian Pisano}
\email{pisano@harpo.physik.uni-dortmund.de}
\affiliation{ Institut f\"ur Physik, Universit\"at Dortmund, D 44221
Dortmund, Germany}
\date{\today\\[2cm]}

\begin{abstract}
The accuracy of the equivalent photon approximation (EPA) of 
the proton in describing the inelastic process $ep\rightarrow \nu W X$ is
investigated. In particular, the scale dependence of the corresponding 
inelastic photon
distribution is discussed. Furthermore, an estimate of the total number 
of events, including the ones coming from the elastic and quasi-elastic 
channels of the reaction, is given for the HERA collider.       
\end{abstract}
\maketitle

\section{Introduction}
The equivalent photon approximation (EPA) of the nucleon $N$ (= $p$, $n$) 
is a technical device which allows for a simpler and more efficient
calculation of any photon-induced subprocess, whose cross section can be 
written as a convolution of the probability that the nucleon radiates off 
a photon (equivalent photon distribution) with the corresponding real 
photoproduction cross section. 
The polarized and unpolarized photon distributions of the nucleon, 
evaluated in the EPA, have been  computed theoretically \cite{gpr1} and 
the possibility of their experimental determination has also  been 
demonstrated 
\cite{pap1,ingo,pp2,pol}. Both of them consist of two components, 
an elastic one, 
due to $N\rightarrow \gamma N $, and an inelastic one, due to $N\rightarrow \gamma X$, with $X \neq N$. 
The reliability of the EPA remains, however,  to be studied.

In \cite{kniehl} the unpolarized elastic photon 
distribution was tested in the case of $\nu W$ production in the process 
$ep\rightarrow \nu W  p$. The relative error of the cross section as 
calculated in the EPA with respect to the exact result
was shown as a function of $\sqrt S$, in the range  
$100 \le \sqrt S \le 1800$ $\mathrm{GeV}$. The agreement turned out to be very 
good, the approximation reproducing the exact cross section within less 
than one percent. Motivated by this results, our aim here is to check if the
same holds in the inelastic channel. 

The process $ep \rightarrow \nu W X $ has been widely studied by several 
authors \cite{alt,gab,baur,neufeld,bohm}. Its relevance is related to the 
possibility of measuring
the three-vector-boson coupling $W W \gamma$, which is a manifestation 
of the nonabelian gauge symmetry upon which the Standard Model is based.
The observation of the vector boson self interaction would be a crucial test
of the theory. Furthermore, such a reaction is also an important background 
to a number of processes indicating the presence of new physics. 
The lightest Supersymmetric Standard Model particle has no charge and 
interacts very weakly with matter; it means that, exactly as the neutrino 
from the Standard Model, it escapes the detector unobserved and can be 
recognized only by missing momentum. This implies that a detailed study of the 
processes with neutrinos in the final states is necessary to distinguish 
between the new physics of the Supersymmetric Standard Model and the physics 
of the Standard Model. 
At the HERA collider energies ($\sqrt S = 318 $ 
GeV) the $e p \rightarrow \nu W  X $ cross section is 
much smaller than the one for $ep \rightarrow e W X $ \cite{gab,baur},
 also sensitive to the $WW\gamma$ coupling, due to the presence in the latter
of an additional 
Feynman graph  where an almost real photon  and a massless quark are 
exchanged in a $u$-channel configuration ($u$-channel pole). The 
dominance of the  process $ep\rightarrow e W X $   justifies the 
higher theoretical and experimental \cite{exp} attention that it has 
received so far, as compared to $ep\rightarrow \nu W X$. One way of limiting 
the problem of the low number of $\nu W$ 
events at HERA would be to consider also the elastic and quasi-elastic 
 channels of the reaction, as will be discussed in 
Section III.  

It is worth mentioning that not all the calculations of  the $ep \rightarrow \nu W X$ event rates available in the literature, in which only the photon
exchange is considered (see Fig. 1), are in agreement, as already pointed out in \cite{bohm}. In particular, the numerical
estimate of the cross section for HERA energies presented   in 
\cite{alt,gab}, obtained in the EPA approach, is one half of the one 
published in \cite{bohm}, obtained within the framework of the helicity 
amplitude
formalism without any approximation.  The value given in \cite{neufeld}
is even bigger than the one in \cite{bohm}: all these discrepancies cannot be
due to the slightly different kinematical cuts employed in the papers cited
above and stimulate a further analysis. Our results agree with \cite{alt,gab}. 

The plan of this paper is as follows. In  Section II we calculate the exact 
cross section for the inelastic channel in a manifestly covariant way and we 
show in which kinematical region it is supposed to be well described by 
the EPA. The formulae for the corresponding elastic cross sections, both
the exact and the one evaluated in the EPA, are also given. 
The numerical results are discussed in Section III. The summary is given in 
Section IV.

\section{Theoretical Framework}

The $\nu W $ production from inelastic $e p $ scattering, 
\be
 e(l)+ p(P) \rightarrow \nu(l')+ W(k')+ X(P_X),
\ee
is described, considering only one photon exchange, by the Feynman 
diagrams depicted in Fig. 1. The four-momenta of the particles are given
in  the brackets; $P_{X}=\sum_{X_i} P_{X_i}$ is the sum over all momenta 
of the produced hadronic system. We introduce the invariants 
\be
S=(P+l)^2, ~~~~~~~~~\hat s=(l+k)^2, ~~~~~~~~Q^2 = -k^2,
\label{invar}
\ee
where $k = P-P_X$ is the four-momentum of the virtual photon. Following 
\cite{pap1}, the integrated cross section can be written as
\be
\sigma_{\mathrm{inel}}(S)&=&{\alpha\over 4 \pi (S-m^2)^2} \int_{W^2_{{\mathrm
{min}}}}^{W^2_{\mathrm{max}}}
 d W^2 \int_{\hat s_{\mathrm{min}}}^{(\sqrt
S-W)^2} d\hat s \int_{Q^2_{\mathrm{min}}}^{Q^2_{\mathrm{max}}} {dQ^2 \over 
Q^4} \int_{\hat t_{\mathrm{min}}}^{\hat
t_{max}} d \hat t \int_0^{2 \pi} d \varphi^* 
\bigg \{ \bigg [\bigg ( 2 \,{{S-m^2}\over {\hat s+Q^2}}\nonumber\\&&~
\times \bigg (1-{{S-m^2}\over {\hat s+Q^2}}\bigg )+(W^2-m^2) \bigg ( {2\,
(S-m^2)\over {Q^2 (\hat s + Q^2)}}-{1\over Q^2} +{m^2-W^2\over 2 \,Q^4}\bigg ) 
\bigg )
\nonumber\\&&~~~~\times [3
X_1(\hat s,Q^2,\hat t)+X_2(\hat s,Q^2,\hat t)]+\bigg ({1\over
Q^2}(W^2-m^2)+{(W^2-m^2)^2\over 2\, Q^4}+{2 m^2\over Q^2} \bigg  )
\nonumber\\&&~~~~~~~\times [X_1(\hat s,Q^2,\hat t)+X_2(\hat s,Q^2,\hat t)]-X_1(\hat
s,Q^2,\hat t)\bigg ] F_2(x_B,Q^2) {x_B\over
2}\nonumber\\&&~~~~~~~~~~~~~~~~~~~~~~~~~~~~~~~~~~-X_2(\hat s,Q^2,\hat
t)F_1(x_B,Q^2)\bigg
\},
\label{siin}
\ee   
where $W^2$ indicates the  invariant mass squared of the 
produced hadronic system $X$, $\varphi^*$ denotes the azimuthal angle of the
outgoing $\nu-W$ system in the $\nu-W$ CM frame, and
\be
x_B = \frac{Q^2}{W^2 + Q^2 - m^2} 
\ee
is the Bjorken variable. 
$F_{1,2}(x_B, Q^2)$ are the structure functions of the proton and 
the two invariants $X_{1,2}(\hat s, Q^2, t)$, which  contain all the 
information about the subprocess $e\gamma^* \rightarrow \nu W$, are given
by 
\be   
   X_1 (\hat s, Q^2, \hat t)& = & \frac{\alpha G_F}{2 \sqrt 2 \pi}
\frac{Q^2 M_W^2}{ (Q^2 + \hat s)^3 \,(M_W^2 - \hat t)^2}   
          [ (Q^2 + \hat s)^3 -\hat s (Q^2 + \hat s)^2 (Q^2 + \hat s + \hat t)  \nonumber \\
&& ~~~~~~ +~ 2 (Q^2 + \hat s)^2 \hat t + 8
          (Q^2 + \hat s) \hat t^2 + 8 \hat t ^3]
\label{xoneex}
\ee
and
\be
 X_2(\hat s , Q^2, \hat t) & = & \frac{\alpha G_F}{2\sqrt 2 \pi}\frac{1}{\hat s ^2 (Q^2 + \hat s )(M_W^2 - \hat t)^2}  \{ 4 M_W^8 (Q^2 + \hat s) - 4 M_W^6 [3 \hat s (Q^2 + \hat s) \nonumber \\ && ~~~+  ~(2 Q^2 + \hat s) \hat t] 
+ 4 M_W^4 [\hat s (2 \hat s + \hat t)^2 + Q^2 (4 \hat s^2 +2 \hat s \hat t + \hat t^2)] - M_W^2 \hat s \nonumber \\&& ~~~~~~~\times~ [Q^4 \hat s + Q^2 (9 \hat s^2    + 2 \hat s \hat t  - 4 \hat t^2) 
+  4 (\hat s + \hat t ) (2 \hat s^2 + 2 \hat s \hat t + \hat t^2)] \nonumber \\&&~~~~~~~~~~~~~~~~+ Q^2 \hat s ^2 [\hat s (\hat s  + \hat t)
 +~ Q^2 (\hat s + 2 \hat t)]  \}.
\label{xtwoex}
\ee
In Eq. (\ref{siin}) the minimum value of $\hat s$ is given by the squared 
mass of the $W$ boson:
\be
\hat s_{\mathrm{min}} = M_W^2,
\label{smin}
\ee 
while the limits of the integration over $W^2$ are:
\be
W^2_{\mathrm{min}}=(m+m_{\pi})^2,~~~~~~~~W^2_{\mathrm{max}}=
(\sqrt S-\sqrt {\hat s_{\mathrm{min}}}\, )^2,
\ee
where $m_{\pi}$ is the mass of the pion.
The limits $Q^2_{\mathrm{min},\mathrm{max}}$ are given by:
\be
Q^2_{{{\mathrm{min}},{\mathrm{max}}}}& = & -m^2-W^2+{1\over 2 S} \Big 
[(S+m^2) (S-\hat s+W^2) \nonumber \\ 
&& ~~~~~~\mp (S-m^2){\sqrt{(S-\hat s+W^2)^2-4 S W^2}} \Big ],
\label{q2lim}
\ee  
and the extrema of $\hat t$ are \be
\hat t_{\mathrm{max}}=0, ~~~~\hat t_{\mathrm{min}} = -\frac{(\hat s + Q^2) (\hat s -M_W^2)}{\hat s}.
\label{thatlim}
\ee
Integrating $X_{1,2} (\hat s, Q^2, \hat t)$ over $\varphi^*$ and 
$\hat t$, with the limits 
in Eq. (\ref{thatlim}), one recovers Eqs. (4.1) and (4.2) of \cite{kniehl} 
respectively, times a factor of two due to a different normalization.   \\
The EPA consists of considering the exchanged photon as real; it is 
possible to get the approximated cross section $\sigma_{\mathrm{inel}}^{\mathrm{EPA}}$ from the exact one, Eq. (\ref{siin}),   in a 
straightforward way, following again {\cite{pap1}}.  We neglect $m^2$ 
compared to $S$ and $Q^2$ compared to $\hat s$
then, from Eqs. (\ref{xone})-(\ref{xtwo}),
\be
{X}_1(\hat s, Q^2,\hat t) \approx {X}_1(\hat s,0, \hat t)=0,
\label{xone}
\ee
and \be
{X}_2(\hat s, Q^2, \hat t)\approx {X}_2(\hat s,0, \hat t) = -{{2 \hat s}\over{\pi}}\, {{d\hat{\sigma}(\hat s, \hat t)}\over{d\hat t}},
\label{xtwo}
\ee
where we have introduced the differential cross section for the real 
photoproduction process $e \gamma \rightarrow \nu W$:
\be
{d \hat \sigma (\hat s, \hat t)\over d\hat t}=-\frac{ \alpha G_F M_W^2}{ \sqrt 2 \hat s^2} \bigg (1 - \frac{1}{1 + \hat u /\hat s} \bigg )^2 ~ \frac{\hat s^2 + \hat u^2 + 2 \hat t M_W^2}{\hat s \hat u}
\label{realsi} 
\ee
with $\hat u = (l-k')^2 =  M_W^2 -\hat s -\hat t$. Eq. (\ref{realsi}) agrees
with the analytical result already presented in \cite{alt,gab}, obtained 
using  the helicity amplitude technique.  
Using Eqs. (\ref{xone}) and ({\ref{xtwo}), we can write
\be
\sigma_{\mathrm{inel}}(S) \approx \sigma_{\mathrm{inel}}^{\mathrm{EPA}} =
\int_{x_{\mathrm{min}}}^{(1-m/\sqrt S)^2} dx \, \int_{M_W^2-\hat s}^0
d\hat t~
\gamma_{\mathrm{inel}}(x, x S) \,{{d\hat\sigma(x S, \hat t)}\over{d\hat t}},
\label{epain}
\ee
where  $x={\hat s/S}$  and $\gamma_{\mathrm{inel}}(x, x S)$ is the 
inelastic component of the equivalent photon distribution of the proton:
\be
\gamma_{\mathrm{inel}} (x, x S)&=&{\alpha\over 2 \pi} \int_x^1\, dy 
\int_{Q^2_{\mathrm{min}}}^{Q^2_{\mathrm{max}}} {dQ^2\over Q^2}\,{y\over x}  
 \bigg [F_2\bigg ({x\over y},Q^2\bigg )\bigg ({{1+(1-y)^2}\over y^2} -
{2 m^2 x^2\over y^2 Q^2} 
\bigg )\nonumber\\&&~~~~~~~~~~~~~~-F_L \bigg({x\over y},Q^2 \bigg) \bigg ],
\label{gammain}
\ee
with
\be
Q^2_{\mathrm{min}}={x^2 m^2\over 1-x}, ~~~~~~Q^2_{\mathrm{max}}= \hat s.
\ee
As pointed out in \cite{alt}, there is some ambiguity in the choice of 
$Q^2_{\mathrm{max}}$, which is typical of all leading logarithmic 
approximations, and any other quantity of the same order of magnitude of 
$\hat s$, like $-\hat t$ or $-\hat u$, would be equally acceptable for 
$Q^2_{\mathrm{max}}$ within the limits of the EPA. The numerical effects
related to the scale dependence of the inelastic photon distribution are 
discussed in the next section.

The cross section relative to the elastic
channel, $ ep \rightarrow \nu W p$, has been calculated in \cite{kniehl} and 
can be written in the form \cite{pap1} 
\be
\sigma_{\mathrm{el}}(S)&=&{\alpha\over 8 \pi (S-m^2)^2} \int_{\hat s_{\mathrm
{min}}}^{(\sqrt
S-m)^2} d\hat s \int_{t_{\mathrm{min}}}^{t_{\mathrm{max}}}{dt \over t} \int_
{\hat t_{\mathrm{min}}}^{\hat t_{{\mathrm{max}}}} d \hat t \int_0^{2 \pi} d \varphi
^* \bigg\{ \bigg [ 2\, {S-m^2\over \hat s-t}
\bigg ( {S-m^2\over \hat s-t}-1 \bigg ) \nonumber\\&& ~~~\times [3 X_1(\hat s,t,\hat
t)+X_2 (\hat s,t,\hat t)] +{2 m^2\over t} [X_1(\hat s,t,\hat t)+X_2(\hat
s,t,\hat t)]+X_1(\hat s,t,\hat t)\bigg ]
H_1(t)\nonumber\\&&~~~~~~~~~~~~~~~~~~~~+X_2(\hat s,t,\hat t) H_2(t) \bigg \},
\label{sigel}
\ee
with $t = -Q^2$, integrated over the range already defined by Eq. 
(\ref{q2lim}),  and $\hat s_{\mathrm{min}}$ given by Eq. (\ref{smin}). 
The limits of integration of $\hat t $ are the same
as  in Eq. (\ref{thatlim}) and the invariants $H_{1,2}(t)$ can be expressed as
\be
H_1(t)={{G_E^2(t)- (t/4 \,m^2)\, G_M^2(t)}\over{1-{t/4\, m^2}}},~~~~~~~~ H_2(t)=G_M^2(t),
\ee
$G_E(t)$ and  $G_M(t)$ being the well-known electric and magnetic form 
factors of the proton, respectively. Again, in the limit $S \gg m^2$ and 
$\hat s \gg -t $, the cross section factorizes and is given by
\be
\sigma_{\mathrm{el}}(S) \approx \sigma_{\mathrm{el}}^{\mathrm{EPA}} =
\int_{x_{\mathrm{min}}}^{(1-{m/ 
\sqrt S})^2}\, dx \,\int_{M_W^2 -\hat s}^0
d\hat t  \,\gamma_{\mathrm{el}} (x) \,{d \hat \sigma (x S, \hat t)\over d\hat t} ,
\label{epael}
\ee
where $x={\hat s/ S}$ and 
\be
\gamma_{\mathrm{el}}(x) =-{\alpha\over 2 \pi} x \int_{t_{\mathrm{min}}}^{t_{\mathrm
{max}}} {dt\over t} \bigg \{ 2 \bigg [ {1 \over x}\bigg  ( {1\over x}-1 \bigg ) 
+{m^2\over t}\bigg ] H_1(t) + H_2(t) \bigg \},
\ee
with 
\be
{t_{\mathrm{min}} \approx -\infty} ~~~~~~~~~~~~t_{\mathrm{max}} \approx 
-{{m^2 x^2}\over{1-x}},
\ee
 is the universal, scale independent, elastic component of the photon 
distribution of the proton, derived for the first time in \cite{kniehl}. 

\section{Numerical Results}
In this section, we present a numerical estimate of the cross sections 
for the reactions $e p \rightarrow \nu W X$ and $ep \rightarrow \nu W p$, 
calculated both exactly and in the EPA, in the range 
$ 100 \le  \sqrt S \le 2000 $ GeV.   We take $M_W = 80.42$ GeV for the 
mass of the $W$ boson and $G_F = 1.1664 \times 10^{-5} $ $\mathrm{GeV}^{-2}$ 
for the Fermi coupling constant \cite{data}. All the integrations 
are performed numerically. In the evaluation of Eqs. (\ref{siin}) and  
(\ref{gammain}) we assume the  LO Callan-Gross relation
\be
F_L(x_B,Q^2)~ =~ F_2(x_B,Q^2)-2 x_B F_1(x_B,Q^2)~=~0,
\ee
and we use the ALLM97 parametrization of the proton structure function 
$F_2(x, Q^2)$ \cite{allm97}, which provides a purely phenomenological, 
Regge model inspired, description of $F_2(x, Q^2)$, including its vanishing 
in the $Q^2 = 0$ limit as well as its scaling behaviour at large $Q^2$. 
The ALLM97 parametrization is supposed to hold over the entire 
range of $x_B$ and $Q^2$ studied so far, namely 
$3 \times 10^{-6} < x_B < 0.85 $ 
and $ 0 \le Q^2 < 5000$ $\mathrm{GeV}^2$, above the quasi-elastic 
region ($W^2 > 3$ 
$\mathrm{GeV}^2$) dominated by resonances. 
We do not consider the resonance contribution 
separately but, using the so-called local duality \cite{bloom}, 
we extend the ALLM97 
parametrization from the continuous ($W^2 > 3$ $\mathrm{GeV}^2$) down to
the resonance  domain ($(m_{\pi}+ m)^2<W^2< 3$ $\mathrm{GeV}^2$): 
in this way it is possible to agree with the experimental data averaged 
over each resonance. In our analysis, the average value of $x_B$ always lies
within the kinematical region mentioned above,  where the experimental 
data are available. 
On the contrary, the avarage value of $Q^2$  becomes larger
than $5000$ $\mathrm{GeV}^2$ when $\sqrt{S} \gtrsim 1200$ GeV, so we need to
extrapolate the  ALLM97 parametrization  beyond the region where the
data have been fitted.   
Our conclusions do not change if we utilize
a parametrization of $F_2(x_B, Q^2)$ whose behaviour at 
large $Q^2$ is constrained by the Altarelli-Parisi evolution
equations, like GRV98  \cite{grv98}.

The electric and magnetic form factors, necessary 
for the determination of the elastic cross sections in Eqs. (\ref{sigel})
and (\ref{epael}),
 are empirically parametrized as dipoles: 
\be
G_E(t)= {1\over [1 - t/(0.71\,\mathrm{GeV}^2)]^{2}},~~~~G_M(t)=2.79~G_E(t).
\ee
At the HERA collider, where the electron and the proton beams 
have energy $E_e = 27.5 $ GeV and $E_p = 920$ GeV respectively, 
the cross section is dominated by the inelastic channel: 
$\sigma_{\mathrm{el}} = 2.47 \times 10^{-2} $ pb, while 
$\sigma_{\mathrm{inel}} = 3.22 \times 10^{-2} $ pb; therefore the expected 
integrated  luminosity of $200$ $\mathrm{pb}^{-1}$ would yield  a total 
of about 11 events/year.
    
Fig. 2  shows a comparison of the inelastic cross section calculated
in the EPA, $\sigma^{\mathrm{EPA}}_{\mathrm{inel}}$, with the exact one, 
$\sigma_{\mathrm{inel}}$, as a function of $\sqrt S$, where several scales for 
$\sigma^{\mathrm{EPA}}_{\mathrm{inel}}$ are proposed, namely
$Q^2_{\mathrm{max}} = \hat s $,\, $-\hat u $, $-\hat t $ in Eq. 
(\ref{gammain})  . It turns out 
that the choice of $-\hat t$ does not provide an adequate description
of $\sigma_{\mathrm{inel}}$, while $\hat s$ and $-\hat u$ are 
approximatively equivalent in reproducing $\sigma_{\mathrm{inel}}$.  
In particular, the choice of
 $- \hat u $ is slightly better in the  range
$ 300 \lesssim  \sqrt S \lesssim 1000$ GeV, while $\hat s $ guarantees 
a more accurate description of the exact cross section for 
$\sqrt S \gtrsim  1000$ GeV.  At HERA energies, 
$\sigma^{\mathrm{EPA}}_{\mathrm{inel}} = 3.64 \times 10^{-2}$ pb,
$3.51 \times 10^{-2}$ pb and
$3.07 \times 10^{-2}$ pb for $Q^2_{\mathrm{max}} = \hat s$, $-\hat u$ and
$-\hat t $, respectively.
In the following we will fix the scale to be 
$\hat s$, in analogy to our previous studies about the QED Compton 
scattering process in $ep \rightarrow e \gamma X$ \cite{pap1,pp2,pol}.  
In \cite{pp2,pol} it was suggested that the experimental selection of only
those events for which $\hat s > Q^2$ restricts the kinematics of the
process to  the region of validity of the EPA and 
improves the  extraction the equivalent photon distribution from the 
exact cross section. The effect of such a cut on the reaction 
$ep \rightarrow \nu W X $ is shown in Fig. 3 and the reduction of the 
discrepancy is evident at large $\sqrt S$, but   not
at  HERA energies, where  $\sigma_{\mathrm{el}} $ and 
$\sigma_{\mathrm{inel}} $ are unchanged. 

In Fig. 4 the total (elastic + inelastic) exact cross section is depicted
as a function of $\sqrt S$, together with the approximated one. 
Here the kinematical constraint $\hat s > Q^2$ is 
{\it not} imposed on the exact cross section.  The average discrepancy is 
reduced to be about $2 \%$, due to the inclusion
of the elastic channel, better described by the EPA (average 
discrepancy $ 0.05  \% $). 
The elastic component is also shown separately, and it agrees with 
the curve presented in Fig. 3 of \cite{kniehl}. For $\sqrt S = 318$ GeV,
$\sigma^{\mathrm{EPA}}_{\mathrm{el}} = 2.47 \times 10^{-2}$ pb, in
perfect agreement with the exact value $\sigma_{\mathrm{el}}$. 

We compare now our results with the ones already published. In \cite{bohm},
taking into account the photon exchange only (Fig. 1) and with no further 
approximation, fixing $M_W = 83.0$ GeV, 
$\sin ^2\theta_W = 0.217$, $E_e = 30$ GeV, 
$E_p = 820$ GeV and using the parton distributions \cite{owens} (Set 1),
together with the cuts $Q^2 > 4$  $\mathrm{GeV}^2$ and  $ W^2> 10$ 
$\mathrm{GeV}^2$, the value $\sigma_{\mathrm{inel}} = 3.0 \times 10^{-2}$ pb  
was obtained. This is  in contrast to 
$\sigma_{\mathrm{inel}} = 1.5 \times 10^{-2}$ pb, calculated using Eq. 
(\ref{siin}) with the same sets of cuts, values of the energies, $M_W$ and 
parton distributions utilized in \cite{bohm}. The authors of  \cite{bohm} 
also report the value $\sigma_{\mathrm{inel}} = 4.0 \times 10^{-2}$ pb, 
obtained in \cite{neufeld} with a similar analysis at the same energies,
using $M_W = 78$ GeV and $\sin ^2\theta_W = 0.217$. The lower limit on $Q^2$  
was taken to be $O(1)$ $\mathrm{GeV}^2$, but not explicitly mentioned. 
Even with the ALLM97 parametrization, which allows us to use no cutoff
on $Q^2$,  we get  $\sigma_{\mathrm{inel}} = 3.1 \times 10^{-2}$ pb, 
far below  $4.0 \times 10^{-2}$ pb.  No analytical 
expression of the cross section is provided in \cite{neufeld,bohm}, which 
makes it difficult to understand the source of the discrepancies. \\
Finally, an estimate of the the 
$ep \rightarrow \nu W X $ cross section is also given  in \cite{alt,gab}, 
utilizing an inelastic equivalent photon distribution   
slightly different
from  the one in Eq. (\ref{gammain}), which can be written in the form
\be
\tilde{\gamma}_{\mathrm{inel}} (x, Q^2_{\mathrm{max}}) = \frac{\alpha}{2\pi} \int_x^1 
d y\, F_2\bigg(\frac{x}{y}, \langle Q^2 \rangle \bigg ) 
~\frac{1 + (1-y)^2}{x\,y} ~\log \frac{Q^2_{\mathrm{max}}}{Q^2_{\mathrm{cut}}},
\label{gabepa}
\ee
where
\be
\langle Q^2 \rangle = \frac{Q^2_{\mathrm{max}} - Q^2_{\mathrm{cut}}}{
\log {\frac{Q^2_{\mathrm{max}}}{Q^2_{\mathrm{cut}} }}},
\ee
$ Q^2_{\mathrm{max}} = x_B S -M_W^2$ and $Q^2_{\mathrm{cut}} = 1$ 
$\mathrm{GeV}^2$. Eq. (\ref{gabepa})
can be obtained from Eq. (\ref{gammain}) neglecting the mass term and 
approximating the integration over $Q^2$. In the 
calculation performed in \cite{gab,alt},  $\tilde{\gamma}_{\mathrm{inel}} 
(x, Q^2_{\mathrm{max}})$ is convoluted with the differential cross section
for the real photoproduction process in Eq. (\ref{realsi}).  
At $\sqrt S = 300$ GeV, fixing $M_W = 84$ GeV,  $\sin ^2\theta_W = 0.217$
and using the parton distribution parametrization 
\cite{owens} (Set 1), we get $\sigma_{\mathrm{inel}} = 1.6 \times 10^{-2}$ pb, 
very close to the value $ 1.5 \times 10^{-2}$ pb published in \cite{alt,gab}.

\section{Summary and Conclusions}
To summarize, we have calculated the cross section for the inelastic  
process $ep\rightarrow \nu W X$,  both exactly and using the 
equivalent photon approximation (EPA) of the proton, in
order to test its accuracy in the inelastic channnel and complete the 
study  initiated in \cite{kniehl}, limited to the  elastic process 
$ep\rightarrow \nu W p$. The relative error of the approximated result
with respect to the exact one is scale dependent; fixing 
the scale to be $\hat s$, it decreases from about $10 \%$ at HERA energies
down to $0.5 \%$  for $\sqrt S = 1500$ GeV, then it slightly increases up 
to $3\%$ for $\sqrt S = 2000$ GeV.  In conclusion, even if not so 
remarkable as for the elastic channel, in which the deviation is always 
below one percent \cite{kniehl}, the approximation can be considered 
quite satisfactory.
We have  compared our calculations with previous ones in the literature
and found that they are in agreement with \cite{alt,gab}, but
disagree with \cite{neufeld,bohm}. Furthermore,
we have estimated the total number of $\nu W$ events expected at the HERA 
collider, including the elastic  and quasi-elastic channels of the 
reaction. The production rate turns out to be quite small, 
about 11 events/year, 
assuming a luminosity of 200 $\mathrm{pb}^{-1}$, but the process 
could still be detected.

\section{acknowledgements}
We would like to thank E. Reya and M. Gl\"uck for many helpful discussions 
and suggestions, as well as for a critical reading of the manuscript.
Discussions with A. Mukherjee are also acknowledged.
This work has been supported in part by
the 'Bundesministerium f\"ur Bildung und Forschung', Berlin/Bonn.

\newpage\begin{center}
%
\parbox{8cm}{\epsfig{figure=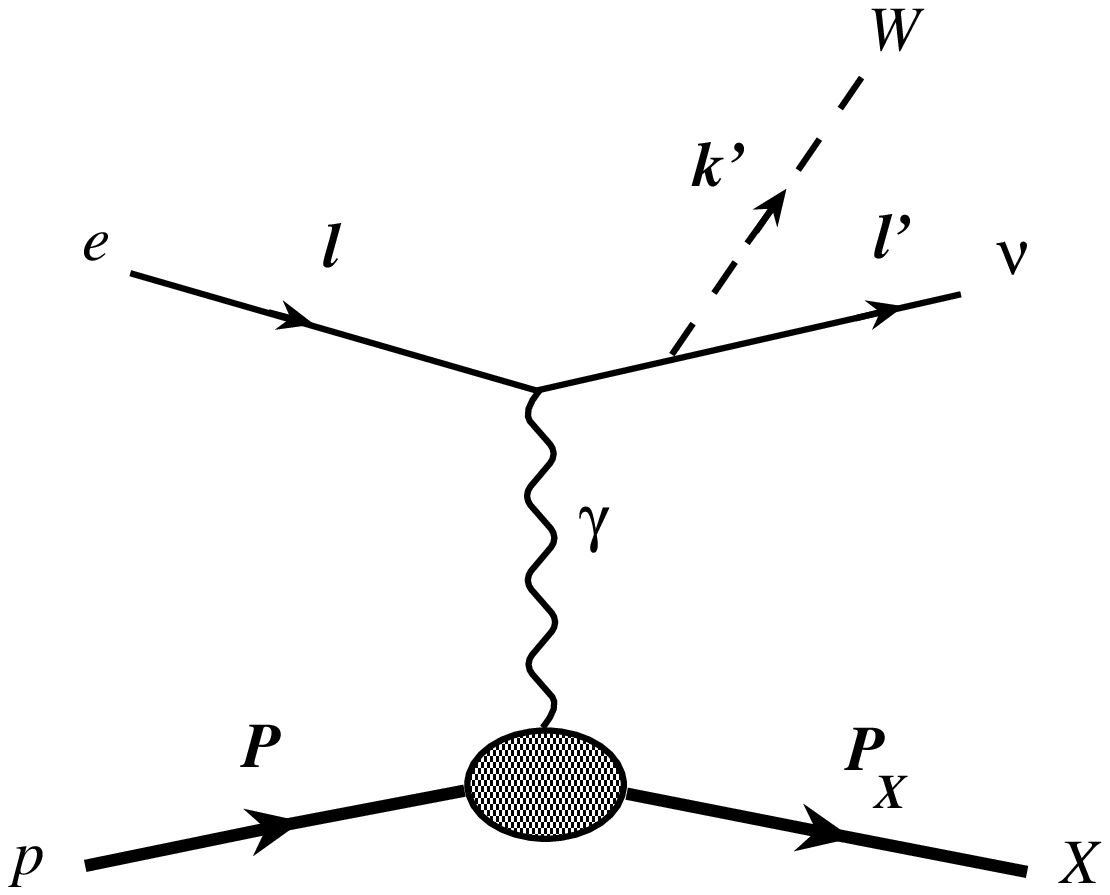,width=8.5cm,height=7.5 cm}}\
\
\parbox{8cm}{\epsfig{figure=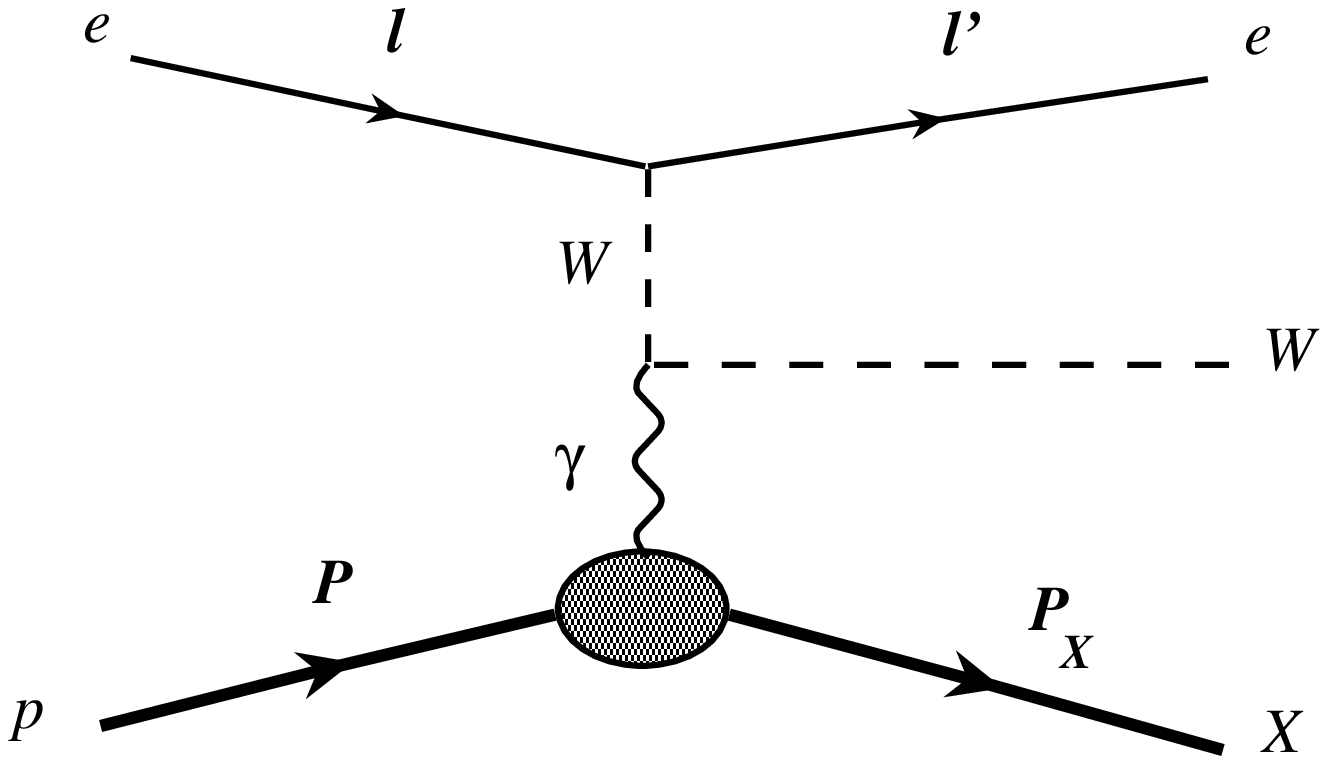,width=8.5cm,height=7.5 cm}}\
\end{center}   
\vspace{0.2cm}
\begin{center}
\parbox{14.0cm}
{{\footnotesize
Fig. 1: Feynman diagrams for the process $ep \rightarrow \nu W X$.}}
\end{center}
  
%

\newpage
\begin{center}
\epsfig{figure=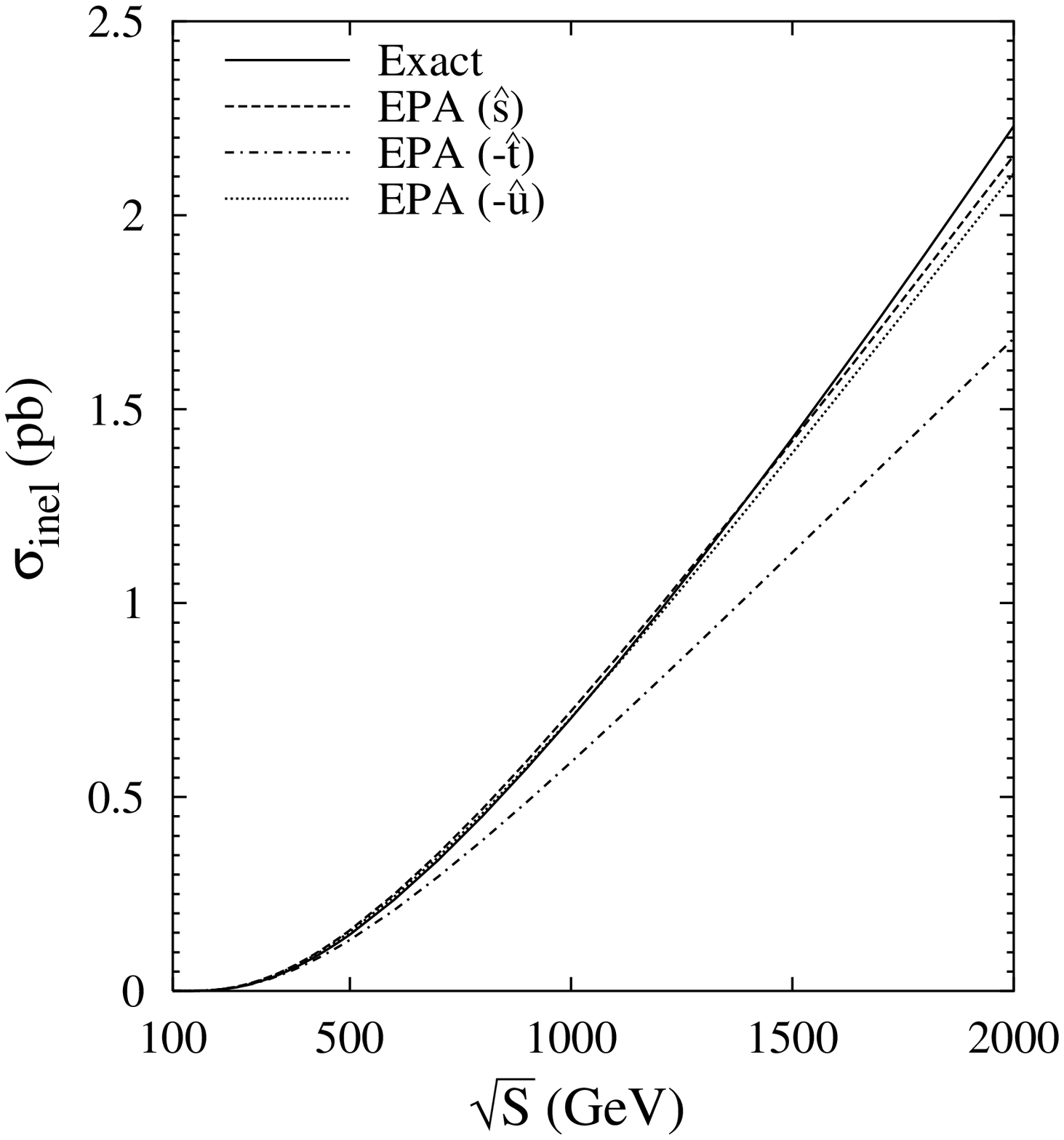, width = 11 cm,height = 10cm}\\
\end{center}   
\vspace{0.2cm}
\begin{center}
\parbox{14.0cm}
{{\footnotesize
Fig. 2: Exact and approximated (EPA) inelastic cross sections of the process 
$ep\rightarrow \nu W X$ as functions of $\sqrt S$. The different scales  
utilized in the calculation of the approximated cross section are written
in the brackets.}}
\end{center}


\newpage
\begin{center}
\epsfig{figure=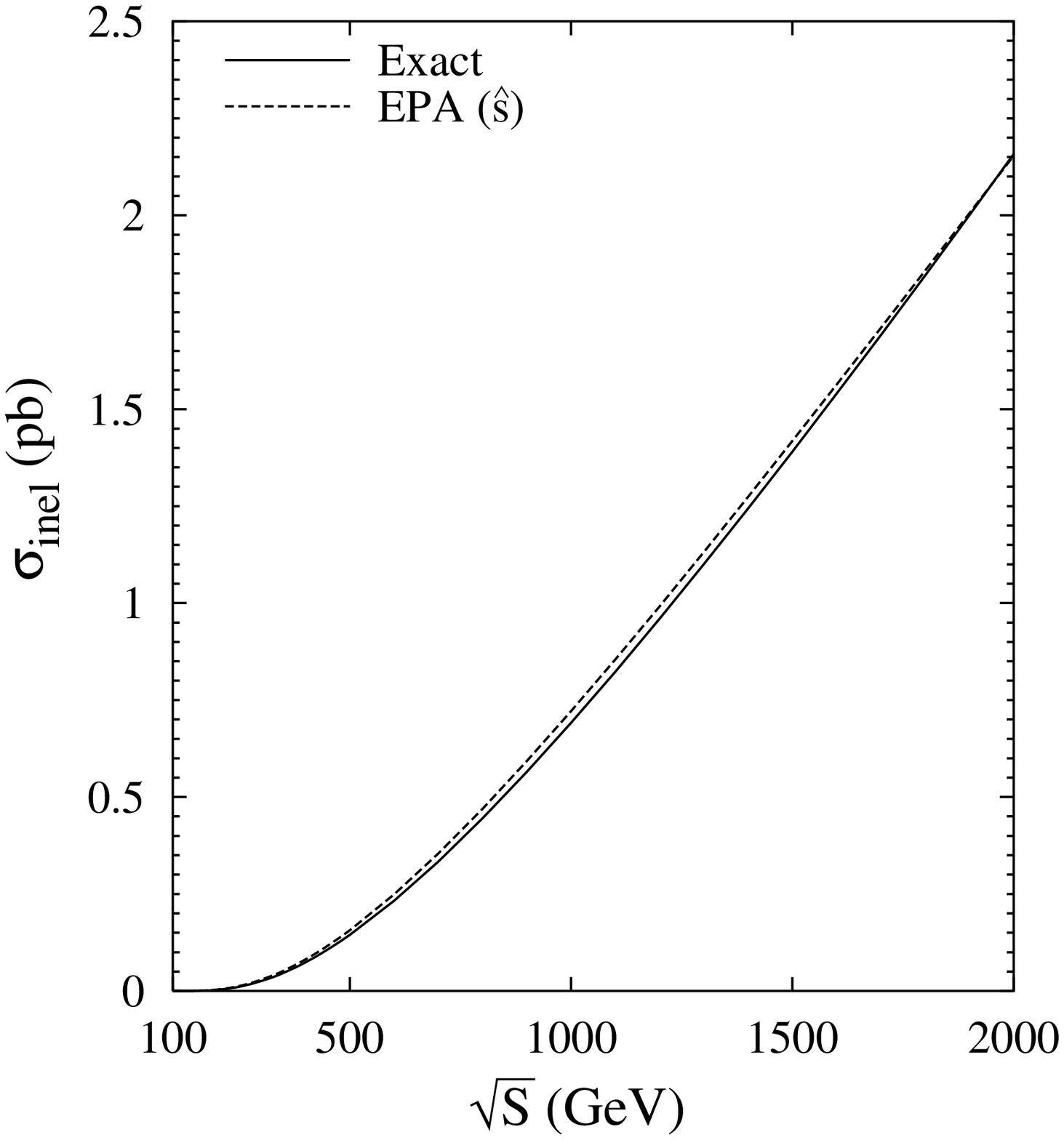, width = 11 cm,height = 10cm}\\
\end{center}   
\vspace{0.2cm}
\begin{center}
\parbox{14.0cm}
{{\footnotesize
Fig. 3: Exact and approximated (EPA) inelastic cross sections of the process 
$ep\rightarrow \nu W X$ as functions of $\sqrt S$. The scale $\hat s$  is 
utilized in the calculation of the approximated cross section and the 
kinematical cut $\hat  s > Q^2 $ is imposed in the exact one.}}
\end{center}


\newpage
\begin{center}
\epsfig{figure=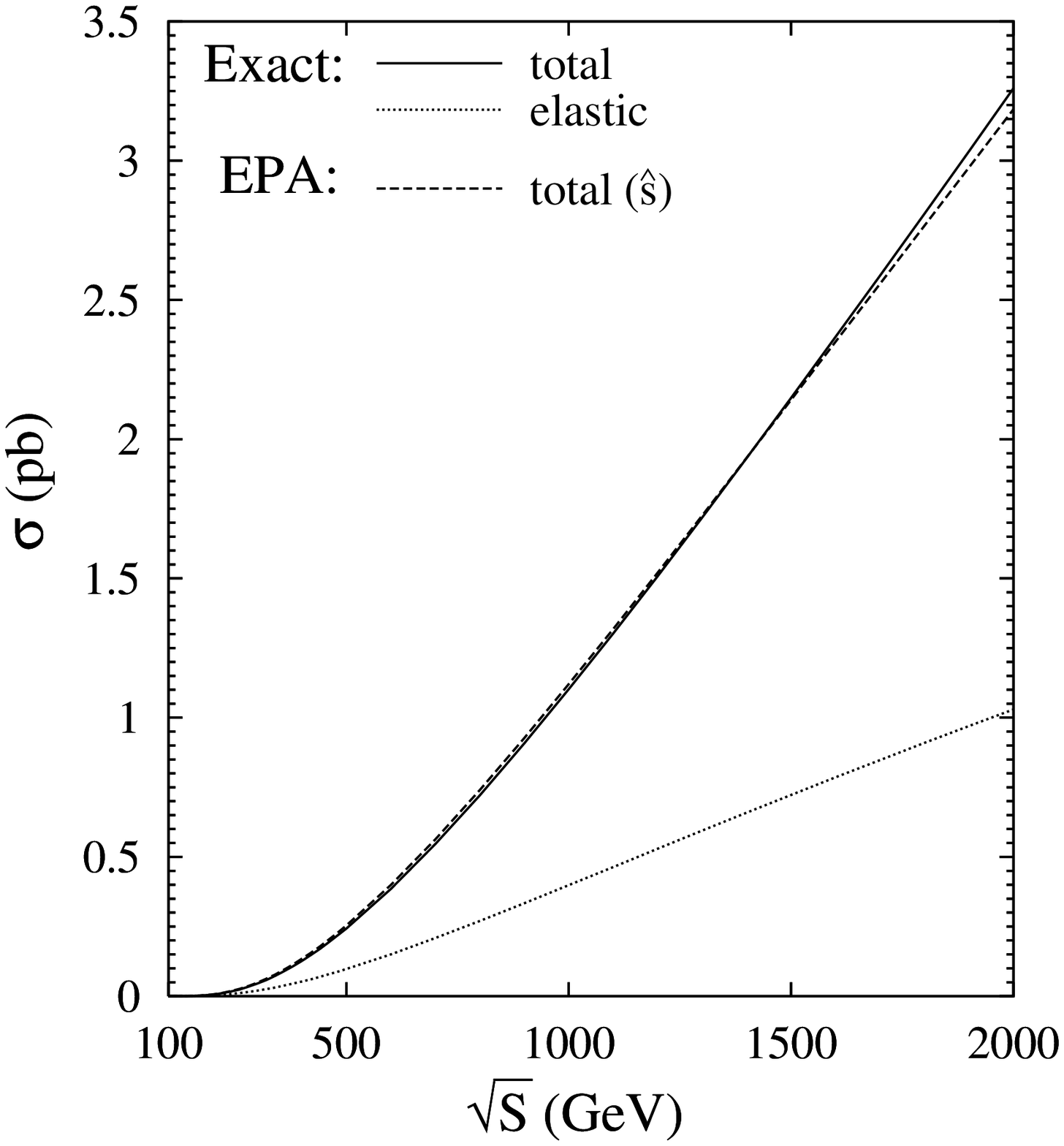, width = 11 cm,height = 10cm}\\
\end{center}   
\vspace{0.2cm}
\begin{center}
\parbox{14.0cm}
{{\footnotesize
Fig. 4: Exact and approximated (EPA) total ( = elastic +  inelastic) cross 
sections of the process 
$ep\rightarrow \nu W X$ as functions of $\sqrt S$.
The exact elastic component, which is indistinguishable from the approximated one,
is shown separately.}}
\end{center}

\end{document}